\documentstyle[12pt,epsfig]{article}

\def\bss{\begin{subequations}}
\def\ess{\end{subequations}}

\catcode`\@=11

\newtoks\@stequation

\def\subequations{\refstepcounter{equation}%
  \edef\@savedequation{\the\c@equation}%
  \@stequation=\expandafter{\theequation}
  \edef\@savedtheequation{\the\@stequation}
  \edef\oldtheequation{\theequation}%
  \setcounter{equation}{0}%
  \def\theequation{\oldtheequation\alph{equation}}}

\def\endsubequations{\setcounter{equation}{\@savedequation}%
  \@stequation=\expandafter{\@savedtheequation}%
  \edef\theequation{\the\@stequation}\global\@ignoretrue
  \vspace*{-12pt} \\}

\catcode`\@=11
\def\@citex[#1]#2{%
\if@filesw \immediate \write \@auxout {\string \citation {#2}}\fi
\@tempcntb\m@ne \let\@h@ld\relax \def\@citea{}%
\@cite{%
  \@for \@citeb:=#2\do {%
    \@ifundefined {b@\@citeb}%
      {\@h@ld\@citea\@tempcntb\m@ne{\bf ?}%
      \@warning {Citation `\@citeb ' on page \thepage \space undefined}}%
      {\@tempcnta\@tempcntb \advance\@tempcnta\@ne%
      \@tempcntb\number\csname b@\@citeb \endcsname \relax%
      \ifnum\@tempcnta=\@tempcntb 
        \ifx\@h@ld\relax%
          \edef \@h@ld{\@citea\csname b@\@citeb\endcsname}%
        \else%
          \edef\@h@ld{\ifmmode{-}\else--\fi\csname b@\@citeb\endcsname}%
        \fi%
      \else
        \@h@ld\@citea\csname b@\@citeb \endcsname%
        \let\@h@ld\relax%
      \fi}%
    \def\@citea{,\penalty\@highpenalty\,}%
  }\@h@ld
}{#1}}

\def\@citeb#1#2{{[#1]\if@tempswa , #2\fi}}
\def\@citeu#1#2{{$^{#1}$\if@tempswa , #2\fi }}
\def\@citep#1#2{{#1\if@tempswa , #2\fi}}

\def\bcites{         
        \catcode`\@=11
        \let\@cite=\@citeb
        \catcode`\@=12
}

\def\upcites{         
        \catcode`\@=11
        \let\@cite=\@citeu
        \catcode`\@=12
}

\def\plaincites{      
        \catcode`\@=11
        \let\@cite=\@citep
        \catcode`\@=12
}

\newcount\hour
\newcount\minute
\newtoks\amorpm
\hour=\time\divide\hour by 60
\minute=\time{\multiply\hour by 60 \global\advance\minute by-\hour}
\edef\standardtime{{\ifnum\hour<12 \global\amorpm={am}%
        \else\global\amorpm={pm}\advance\hour by-12 \fi
        \ifnum\hour=0 \hour=12 \fi
        \number\hour:\ifnum\minute<10 0\fi\number\minute\the\amorpm}}
\edef\militarytime{\number\hour:\ifnum\minute<10 0\fi\number\minute}

\def\draftlabel#1{{\@bsphack\if@filesw {\let\thepage\relax
   \xdef\@gtempa{\write\@auxout{\string
      \newlabel{#1}{{\@currentlabel}{\thepage}}}}}\@gtempa
   \if@nobreak \ifvmode\nobreak\fi\fi\fi\@esphack}
        \gdef\@eqnlabel{#1}}
\def\@eqnlabel{}
\def\@vacuum{}
\def\marginnote#1{}
\def\draftmarginnote#1{\marginpar{\raggedright\scriptsize\tt#1}}
\def\draft2label#1{\draftmarginnote{#1} {\@bsphack\if@filesw {\let\thepage\relax
   \xdef\@gtempa{\write\@auxout{\string
      \newlabel{#1}{{\@currentlabel}{\thepage}}}}}\@gtempa
   \if@nobreak \ifvmode\nobreak\fi\fi\fi\@esphack}
        \gdef\@eqnlabel{#1}}
\def\draft{
        \pagestyle{plain}
        \overfullrule=2pt
        \oddsidemargin -.5truein
        \def\@oddhead{\sl \phantom{\today\quad\militarytime} \hfil
        \smash{\Large\sl DRAFT} \hfil \today\quad\militarytime}
        \let\@evenhead\@oddhead
        \let\label=\draftlabel
        \let\marginnote=\draftmarginnote
        \def\ps@empty{\let\@mkboth\@gobbletwo
        \def\@oddfoot{\hfil \smash{\Large\sl DRAFT} \hfil}
        \let\@evenfoot\@oddhead}
        \def\@eqnnum{(\theequation)\rlap{\kern\marginparsep\tt\@eqnlabel}%
        \global\let\@eqnlabel\@vacuum}  }

\def\draft2{
        \pagestyle{plain}
        \overfullrule=2pt
        \oddsidemargin -.5truein
        \def\@oddhead{\sl \phantom{\today\quad\militarytime} \hfil
        \smash{\Large\sl DRAFT} \hfil \today\quad\militarytime}
        \let\@evenhead\@oddhead
        \let\label=\draft2label
        \let\marginnote=\draftmarginnote
        \def\ps@empty{\let\@mkboth\@gobbletwo
        \def\@oddfoot{\hfil \smash{\Large\sl DRAFT} \hfil}
        \let\@evenfoot\@oddhead} }

\def\blackfonts{
        \font\blackboard=msbm10 scaled\magstep1
        \font\blackboards=msbm8
        \font\blackboardss=msbm6
}

\def\nblack{            
        \def\ZZ{{Z \n{10} Z}}
        \def\NN{{N \n{14} N}}
        \def\CC{{C \n{11} C}}
        \def\RR{{R \n{11} R}}
        \def\QQ{{Q \n{12} Q}}
        \def\PP{{P \n{11} P}}
}

\def\prep{         
        \catcode`\@=11
        \input art10.sty
        \catcode`\@=12
        
        \let\small\null
        \def\blackfonts{
                \font\blackboard=msbm10
                \font\blackboards=msbm7
                \font\blackboardss=msbm5
        }
        \let\sl\it
        \twocolumn
        \sloppy
        \voffset=-2.54truecm
        \hoffset=-2.54truecm
        \flushbottom
        \parindent 1em
        \leftmargini 2em
        \leftmarginv .5em
        \leftmarginvi .5em
        \marginparwidth 48pt
        \marginparsep 10pt
        \setlength{\columnsep}{2truecm}
        \setlength{\textwidth}{25.4truecm}
        \setlength{\textheight}{17truecm}
        \baselineskip=16pt
        \oddsidemargin .18truein
        \evensidemargin .17truein
}

\def\eqalign#1{\null\,\vcenter{\openup\jot\m@th
  \ialign{\strut\hfil$\displaystyle{##}$&$\displaystyle{{}##}$\hfil
      \crcr#1\crcr}}\,}
\def\eqalignno#1{\displ@y \tabskip\centering
  \halign to\displaywidth{\hfil$\@lign\displaystyle{##}$\tabskip\z@skip
    &$\@lign\displaystyle{{}##}$\hfil\tabskip\centering
    &\llap{$\@lign##$}\tabskip\z@skip\crcr
    #1\crcr}}

\def\section{\@startsection {section}{1}{\z@}{3.ex plus 1ex minus
 .2ex}{2.ex plus .2ex}{\large\bf}}
\def\subsection{\@startsection{subsection}{2}{\z@}{2.75ex plus 1ex minus
 .2ex}{1.5ex plus .2ex}{\bf}}        

\def\appendix{{\newpage\section*{Appendix}}\let\appendix\section%
        {\setcounter{section}{0}
        \gdef\thesection{\Alph{section}}}\section}

\def\abstract{\if@twocolumn
\section*{Abstract}
\else 
\begin{center}
{\bf Abstract\vspace{-.5em}\vspace{0pt}}
\end{center}
\quotation
\fi}

\catcode`\@=12

\newcommand{\beq}{\begin{equation}}
\newcommand{\eeq}{\end{equation}}
\newcommand{\beqa}{\begin{eqnarray}}
\newcommand{\eeqa}{\end{eqnarray}}

\def\noj#1,#2,{{\bf #1} (19#2)\ }
\def\jou#1,#2,#3,{{\sl #1\/ }{\bf #2} (19#3)\ }
\def\ann#1,#2,{{\sl Ann.\ Physics\/ }{\bf #1} (19#2)\ }
\def\cmp#1,#2,{{\sl Comm.\ Math.\ Phys.\/ }{\bf #1} (19#2)\ }
\def\ma#1,#2,{{\sl Math.\ Ann.\/ }{\bf #1} (19#2)\ }
\def\ng#1,#2,{{\sl Nagoya.\ Math.\ J.\/ }{\bf #1} (19#2)\ }
\def\jd#1,#2,{{\sl J.\ Diff.\ Geom.\/ }{\bf #1} (19#2)\ }
\def\invm#1,#2,{{\sl Invent.\ Math.\/ }{\bf #1} (19#2)\ }
\def\cq#1,#2,{{\sl Class.\ Quantum Grav.\/ }{\bf #1} (19#2)\ }
\def\cqg#1,#2,{{\sl Class.\ Quantum Grav.\/ }{\bf #1} (19#2)\ }
\def\ijmp#1,#2,{{\sl Int.\ J.\ Mod.\ Phys.\/ }{\bf A#1} (19#2)\ }
\def\jmphy#1,#2,{{\sl J.\ Geom.\ Phys.\/ }{\bf #1} (19#2)\ }
\def\jams#1,#2,{{\sl J.\ Amer.\ Math.\ Soc.\/ }{\bf #1} (19#2)\ }
\def\grg#1,#2,{{\sl Gen.\ Rel.\ Grav.\/ }{\bf #1} (19#2)\ }
\def\mpl#1,#2,{{\sl Mod.\ Phys.\ Lett.\/ }{\bf A#1} (19#2)\ }
\def\nc#1,#2,{{\sl Nuovo Cim.\/ }{\bf #1} (19#2)\ }
\def\np#1,#2,{{\sl Nucl.\ Phys.\/ }{\bf B#1} (19#2)\ }
\def\pl#1,#2,{{\sl Phys.\ Lett.\/ }{\bf #1B} (19#2)\ }
\def\pla#1,#2,{{\sl Phys.\ Lett.\/ }{\bf #1A} (19#2)\ }
\def\pr#1,#2,{{\sl Phys.\ Rev.\/ }{\bf #1} (19#2)\ }
\def\prd#1,#2,{{\sl Phys.\ Rev.\/ }{\bf D#1} (19#2)\ }
\def\prl#1,#2,{{\sl Phys.\ Rev.\ Lett.\/ }{\bf #1} (19#2)\ }
\def\prp#1,#2,{{\sl Phys.\ Rept.\/ }{\bf #1C} (19#2)\ }
\def\ptp#1,#2,{{\sl Prog.\ Theor.\ Phys.\/ }{\bf #1} (19#2)\ }
\def\ptpsup#1,#2,{{\sl Prog.\ Theor.\ Phys.\/ Suppl.\/ }{\bf #1} (19#2)\ }
\def\rmp#1,#2,{{\sl Rev.\ Mod.\ Phys.\/ }{\bf #1} (19#2)\ }
\def\yadfiz#1,#2,#3[#4,#5]{{\sl Yad.\ Fiz.\/ }{\bf #1} (19#2) #3%
\ [{\sl Sov.\ J.\ Nucl.\ Phys.\/ }{\bf #4} (19#2) #5]}
\def\zh#1,#2,#3[#4,#5]{{\sl Zh.\ Exp.\ Theor.\ Fiz.\/ }{\bf #1} (19#2) #3%
\ [{\sl Sov.\ Phys.\ JETP\/ }{\bf #4} (19#2) #5]}

\def\beq{\begin{equation}}
\def\eeq{\end{equation}}
\def\beqar{\begin{eqnarray}}
\def\eeqar{\end{eqnarray}}

\newcommand{\be}{\begin{equation}}
\newcommand{\ee}{\end{equation}}
\newcommand{\bea}{\begin{eqnarray}}
\newcommand{\eea}{\end{eqnarray}}

\def\nfrac#1#2{{\displaystyle{\vphantom1\smash{\lower.5ex\hbox{\small$#1$}}%
        \over\vphantom1\smash{\raise.25ex\hbox{\small$#2$}}}}}

\def\n#1{\mskip-#1mu}

\def\lae{\mathrel{\mathop{\smash{\lower .5 ex \hbox{$\stackrel<\sim$}}}}}
\def\lae{\mathrel{\mathop{\smash{\lower .5 ex \hbox{$\stackrel>\sim$}}}}}

\def\l:{\mathopen{:}\,}
\def\r:{\,\mathclose{:}}

\catcode`\@=11
\def\theequation{\arabic{equation}}
\catcode`\@=12

\nblack
\bcites

\nblack

\catcode`\@=11
\def\theequation{\thesection.\arabic{equation}}
\@addtoreset{equation}{section}
\@addtoreset{footnote}{section}
\@addtoreset{footnote}{subsection}
\catcode`\@=12

\newcommand{\beqn}{\begin{equation}}
\newcommand{\eeqn}{\end{equation}}
\newcommand{\beqnarray}{\begin{eqnarray}}
\newcommand{\eeqnarray}{\end{eqnarray}}
\newcommand {\bear} [1] {\begin {array} {#1}}
\newcommand {\ear} {\end {array}}

\newcommand {\beqarn} {\begin{eqnarray*}}
\newcommand {\eeqarn} {\end{eqnarray*}}

\def\nonu{\nonumber}

\def\diffp{\partial_+}
\def\diffm{\partial_-}
\def\a{\alpha}

\def\s{\sigma}
\def\cl{{\cal L}}

\def\delh{\hat{\nabla}}

\def\delhp{\hat{\nabla}^+}
\def\delhm{\hat{\nabla}^-}

\def\lt{\tilde{L}}
\def\lbt{\tilde{\bar{L}}}

\def\tt{\tilde{T}}
\def\tbt{\tilde{\bar{T}}}

\def\ppf{\frac{1}{4 \pi \a'}}

\begin{document}

\begin{titlepage}

\begin{center}
\today
\hfill LBNL-41142, UCB-PTH-98/01\\
\hfill                  hep-th/9802079

\vskip 1.5 cm
{\large \bf Conformal Field Theories: From Old 
to New}\footnote{To appear in a memorial issue of Theoretical and
Mathematical Physics in memory of F.A. Lunev.}
\vskip 1 cm 
{J. de Boer and M. B. Halpern}\\
\vskip 0.5cm
{\sl Department of Physics,
University of California at Berkeley\\
366 Le\thinspace Conte Hall, Berkeley, CA 94720-7300, U.S.A.\\
and\\
Theoretical Physics Group, Mail Stop 50A--5101\\
Ernest Orlando Lawrence Berkeley National Laboratory\\
Berkeley, CA 94720, U.S.A.\\}

\end{center}

\vskip 0.5 cm
\begin{abstract}

In a short review of recent work, we discuss the general problem of
constructing the actions of new conformal field theories from old
conformal field theories. Such a construction follows when the old conformal
field theory admits new conformal stress tensors in its chiral algebra,
and it turns out that
the new conformal field theory is generically a new spin-two 
gauge theory. As an example we discuss the new
spin-two gauged sigma models
which arise in this fashion from the general conformal non-linear
sigma model.

\end{abstract}

\end{titlepage}

\section{Introduction}

We are saddened by the death of Dr. F.A. Lunev and offer this contribution
in his honor.

The problem of constructing new conformal field theories from old
conformal field theories dates back to $K$-conjugation 
covariance \cite{c1,c2,c3,c4,c5}, the coset
constructions \cite{c1,c2,c3} and the general affine-Virasoro construction
\cite{c5,c6,c7}.
Such a construction follows when the old conformal
field theory admits new conformal stress tensors in its chiral algebra.
The simplest examples of this construction are the coset
constructions, whose (new) spin-one gauged WZW actions \cite{c8,c9,kahs,kas} 
are obtained from the (old) WZW actions. The coset constructions are
however a special case of higher symmetry, and
the problem of finding the actions of the generic affine-Virasoro 
constructions was solved in \cite{c10,c11,c12}, where the (new) spin-two
gauged WZW actions of these theories 
were obtained from the (old) WZW action.

Recently,
we have extended \cite{c13,c14,c15} this program to the new
conformal field theories which can be obtained in this way from
the (old) general non-linear sigma model, and this memorial issue
provides an opportunity to review the program here in general
terms. The discussion of the second and third sections is based 
on material originally discussed in \cite{c10,c16}, and the discussion
of Section~4 is based on material originally discussed in \cite{c13,c14,c15}.

\section{Action Formulation of a CFT}

Consider a conformal field theory (CFT) $C_{\ast}$ with chiral/antichiral
stress tensors $T_{\ast}$, $\bar{T}_{\ast}$ and associated
chiral/antichiral algebras ${\cal A}_{\ast}$, $\bar{\cal A}_{\ast}$
\be
T_{\ast} \in {\cal A}_{\ast}, \qquad \bar{T}_{\ast} \in\bar{\cal A}_{\ast}
\ee
where ${\cal A}_{\ast}$, $\bar{\cal A}_{\ast}$ are defined to include
all mutually local holomorphic/antiholomorphic objects in the theory.
The naive Hamiltonian of $C_{\ast}$ is
\be
H_{\ast 0} = \int_0^{2\pi} d\s\, {\cal H}_{\ast 0},\qquad
{\cal H}_{\ast 0} = T_{\ast} +\bar{T}_{\ast}
\ee
but $C_{\ast}$ is a gauge theory if the centralizers 
${\cal A}'_{\ast}$, $\bar{\cal A}'_{\ast}$ of ${\cal H}_{\ast 0}$ in
${\cal A}_{\ast}$, $\bar{\cal A}_{\ast}$,
\be
{\cal A}'_{\ast} = \{X(z) \in {\cal A}_{\ast} | [X(z),T_{\ast}]=0 \},
\qquad\bar{\cal A}'_{\ast} = \{X(\bar{z}) \in \bar{\cal A}_{\ast}  | 
[X(\bar{z}),\bar{T}_{\ast} ] =0 \}
\ee
are non-trivial\footnote{Usually one defines the chiral algebra
of a conformal field theory such that these centralizers contain
only the unit operator, the gauge degrees 
of freedom having already been modded out.
However, we first allow for a more general situation
where the conformal field theory is embedded in a larger 
gauge-covariant 
system, modding out later by the gauge degrees of freedom.}.
We assume for simplicity that all elements of
the centralizer can be expressed as differential polynomials in terms
of a finite local set of basis elements, which we also denote by
${\cal A}'_{\ast}$, $\bar{\cal A}'_{\ast}$. The centralizers are
in fact infinite dimensional ${\bf Z}$-graded algebras, 
and the corresponding
positive frequency modes of the centralizers will be
denoted by ${\cal A}'_{\ast}(+)$, $\bar{\cal A}'_{\ast}(+)$. 
Following Gupta and Bleuler, 
the theory 
$C_{\ast}$ can then
be described by the Hamiltonian $H_{\ast 0}$
acting on a physical Hilbert space defined by
\be
{\cal A}'_{\ast}(+)|{\rm phys}\rangle =\bar{\cal A}'_{\ast}(+)
|{\rm phys}\rangle =0.
\ee
This means that the physical states are primary under the algebras
${\cal A}'_{\ast}$, $\bar{\cal A}'_{\ast}$.

When $C_{\ast}$ has a smooth classical limit, we expect that the
theory has an action description, To find the action one must first
find the classical limit or Poisson bracket description of the objects
and algebras described above. We assume for simplicity that the 
classical limit of the centralizer contains no central terms, that
is, the basis elements of the centralizer form a set of first-class
constraints in the language of Dirac. Then the full classical 
Hamiltonian of the the CFT $C_{\ast}$ can be written as
\be
H_{\ast} = \int d\s\, {\cal H}_{\ast}, \qquad
{\cal H}_{\ast} = {\cal H}_{\ast 0} + 
v\cdot {\cal A}'_{\ast} + \bar{v} \cdot \bar{\cal A}'_{\ast}
\ee
where $v,\bar{v}$ are Lagrange multipliers, and the action
$S_{\ast}$ of $C_{\ast}$ is obtained by the usual canonical
method. The multipliers $v,\bar{v}$ form world-sheet gauge fields
whose spins are those of the corresponding elements of
the centralizers ${\cal A}'_{\ast},
\bar{\cal A}'_{\ast}$ of $C_{\ast}$.

\section{New CFT's}

We focus now on a case of particular interest, when the 
chiral/antichiral algebras ${\cal A}_{\ast}$, $\bar{\cal A}_{\ast}$  
of $C_{\ast}$ contain two {\it new} chiral/antichiral spin-two
objects $T$ and $\bar{T}$ which satisfy commuting
Virasoro algebras. In this case, one expects the existence of 
two further chiral/antichiral spin-two tensors $\tt$, $\tbt$,
so that all four stess tensors 
\be
T,\tt,\bar{T},\tbt;\qquad
T,\tt \in {\cal A}_{\ast}, \quad 
\bar{T},\tbt \in \bar{\cal A}_{\ast}
\ee
are commuting Virasoro operators. The four new
stress tensors sum in pairs to the stress tensors $T_{\ast}$,
$\bar{T}_{\ast}$ of $C_{\ast}$, 
\be T_{\ast}=T+\tt,\qquad \bar{T}_{\ast}=\bar{T} + \tilde{\bar{T}}
\ee
and this is called $K$-conjugation covariance \cite{c1,c2,c3,c4,c5}.
This phenomenon is most familiar in the coset constructions 
(where $T_{\ast}=T_g$, $T=T_{g/h}$, $\tt=T_h$) and is 
explicit in the general affine-Virasoro construction, but
the phenomenon of $K$-conjugation covariance was 
argued quite generally in \cite{c4}.

In the presence of $K$-conjugation covariance we see that $C_{\ast}$ 
is a tensor product CFT composed of the $K$-conjugate pair of CFTs
$C$ and $\tilde{C}$
\bss
\be
C_{\ast} = C \otimes \tilde{C}
\ee
\be
C:\,T,\bar{T},\qquad \tilde{C}:\tt,\tilde{\bar{T}} 
\label{abc}
\ee
\ess
whose stress tensors are shown in (\ref{abc}).
In what follows we focus on the $C$ theory, but
the corresponding description of the $\tilde{C}$ theory
can be obtained at any stage of the discussion by $K$-conjugation.

The naive Hamiltonian of the $C$ theory is
\be H_0=\int d\s\, {\cal H}_0, \qquad {\cal H}_0=T+\bar{T} \ee
and we see that $C$ is generically\footnote{The case of the $g/h$
coset constructions 
is a special case of higher symmetry: These are spin-one gauge
theories because the centralizer of $T_{g/h}$ is generated by
the $h$-currents, with $T_h$ a composite operator.} a spin-two
gauge theory because the Virasoro operators
$\tt$ and $\tilde{\bar{T}}$ are in the
centralizer ${\cal A}'$, $\bar{\cal A}'$ of ${\cal H}_0$. In the
classical limit, the full Hamiltonian of the generic theory $C$
is therefore
\be H=\int d\s\, {\cal H}, \qquad {\cal H}=
T+\bar{T}+v\tt + \bar{v} \tilde{\bar{T}}
\label{pqr}
\ee
where $v$ and $\bar{v}$ form a spin-two gauge field on the
world-sheet.

Generically, the centralizer of $H_0$ is nothing but $\tt$ and
$\tbt$ so the Hamiltonian (\ref{pqr}) is the proper 
description of the generic new CFT $C$. In special cases of higher
symmetry however, we must gauge the new theory $C$ by
the full centralizer of ${\cal H}_0$,
\be 
\label{xx}
{\cal H}=
T+\bar{T}+v\cdot {\cal A}' + \bar{v} \cdot \bar{{\cal A}}'
\ee
where ${\cal A}'$, $\bar{\cal A}'$ may satisfy Virasoro algebras,
$W_3$ algebras, etc. 
Adding to (\ref{xx}) a 
term proportional to $v\bar{v}$, one can
also include local spin-one symmetries, which are associated to
affine Lie algebras.
The actions of the new CFTs then follow
by the usual canonical prescription.

\section{New Spin-Two Gauged Sigma Models}

We have recently studied this program \cite{c14,c15}
starting with the action of the general conformal
non-linear sigma model 
\be
S_{\ast} = S_G = \int d^2 \xi \, {\cal L}_G, \qquad
{\cal L}_G = (G_{ij}+B_{ij}) \diffp x^i \diffm x^j
\ee
with conformal stress tensors $T_G,\bar{T}_G$. 
Taking the sigma model as a background, we looked 
for all four new chiral/antichiral stress tensors 
 $T,\tt,\bar{T},\tilde{\bar{T}}$ in the generic form
\be
T\sim L_{ij} \diffp x^i \diffp x^j 
\label{w1}
\ee
and found that they exist in $K$-conjugate pairs
\be
T_{\ast}=T_G=T+\tt, \qquad \bar{T}_{\ast}=\bar{T}_G = \bar{T} + \bar{T}_G
\label{w2}
\ee
under certain conditions on the coefficients $L_{ij}$, given
below (see (\ref{q1}) and (\ref{q2})). Including the dilaton,
the quantum extension 
of (\ref{w1}) and (\ref{w2}) has been verified at the one-loop
level in \cite{c13,c14}.

Following the program outlined above, we found that the
generic new conformal field theories $C=T,\bar{T}$ are described 
by the following set of new spin-two gauged sigma models
\bss
\bea
\label{eq52}
S& =& \int d^2 \xi\, \cl
\\{}
\cl & = & \cl_G + \ppf [\alpha \lt_{ij} B^i B^j +
\bar{\alpha} \lbt_{ij} \bar{B}^i \bar{B}^j 
\nonu \\{}
& & \qquad \qquad
 \quad -(B^i-\diffp x^i) G_{ij} (\bar{B}^j - \diffm x^j) ]
\label{eq54}
\eea
\ess
\bss
\bea
& \delhp_i \lt_j{}^k = \delhm_i \lbt_j{}^k =0 & 
\label{q1}
\\{}
& \lt_i{}^j = 2 \lt_i{}^k \lt_k{}^j , \qquad
\lbt_i{}^j = 2 \lbt_i{}^k \lbt_k{}^j . & 
\label{q2}
\eea
\ess
Here $\alpha,\bar{\alpha}$ form a spin-two gauge field, closely related 
to the multipliers $v,\bar{v}$, and $B$, $\bar{B}$ are
auxiliary fields.
One  new $C=T,\bar{T}$
conformal field theory is obtained for each solution
of the conditions in (\ref{q1}), (\ref{q2}). The
gradients $\delh^{\pm}$ in (\ref{q1}) are generalized covariant
derivatives with torsion. See \cite{c15} for further details, including
the spin-two gauge invariance of these actions and their non-linear
form after integrating out the auxiliary fields.

The relations (\ref{q1}), (\ref{q2}) are nothing but the
conditions that the classical chiral algebras are closed,
and necessary and sufficient conditions for their
solution are also discussed in \cite{c13,c14,c15}.
The following explicit examples of this system have been discussed:\\{}
$\bullet\quad$ the spin-two gauged WZW actions\\{}
$\bullet\quad$ the spin-two gauged $g/h$ coset constructions.\\{}
In the first case, the action $S_{\ast}=S_{WZW}$ is the
WZW action and the action $S$ in (\ref{eq52}) is the
generic affine-Virasoro action \cite{c10,c11,c12}, which describes
the generic affine-Virasoro construction. In the second case \cite{c15}, 
where $S_{\ast}=S_{g/h}$ is the sigma model formulation of the coset
constructions, the actions $S$ have
been identified as the actions of the Lie $h$-invariant 
CFTs \cite{c17}, which are those generically irrational CFTs
with an extra local $h$ gauge symmetry. (Because they have an
extra local $h$ symmetry, the Lie $h$-invariant CFTs are not generic and
are therefore
not included in the generic spin-two gauged WZW action.)

In the case that the centralizer of $T$ is larger
than the Virasoro algebra, we can still construct an 
action for the $C=T,\bar{T}$
theory. If the centralizer is generated by holomorphic/antiholomorphic
polynomials $P_r(x^i,\diffp x^i)$, $\bar{P}_r(x^i,\diffm x^i)$,
the action is of the form (\ref{eq54}), 
\bss
\bea
\label{eq2}
S& =& \int d^2 \xi\, \cl
\\{}
\cl & = & \cl_G + \ppf [ \sum_r \alpha_r P_r(x^i,B^i) +
\sum_r \bar{\alpha}_r \bar{P}_r(x^i,\bar{B}^i)
\nonu \\{}
& & \qquad \qquad
 \quad -(B^i-\diffp x^i) G_{ij} (\bar{B}^j - \diffm x^j) ]
\label{eq4}
\eea
\ess
where $\alpha \lt_{ij}
B^i B^j$ has been
replaced by $\sum_r \alpha_r P_r(x^i,B^i)$, and
similarly for the term involving $\bar{\alpha}$. It is quite 
remarkable that by introducing auxiliary fields this large class of
actions can be brought to this simple polynomial form. Integrating 
out the auxiliary fields yields the non-linear
form of these actions, which are also non-local in the general case.

The action (\ref{eq54}) and its generalizations could serve
as a starting point for a perturbative BRST quantization of the new conformal
field theories (see \cite{c11}). 
It would be interesting to study this and the relation
to non-critical string theories in more detail. 

The action (\ref{eq54}) describes the new conformal field theories
in a conformal gauge for the $C$ theory, that is, the world-sheet
metric $h_{mn}$ of the $C$ theory is proportional to $\delta_{mn}$.
By also gauging the stress tensors $T,\bar{T}$ of the $C$ theory
\be
H_2=\int d\sigma\, {\cal H}_2, \qquad
{\cal H}_2 = uT+\bar{u}\bar{T} + v \tt + \bar{v} \tbt
\ee
we obtain a ``doubly-gauged'' action \cite{c10,c11,c12,c15} for the
new CFTs with an arbitrary world-sheet metric $h_{mn}$ for the 
$C$ theory, which is composed of the gauge fields $u,\bar{u}$.
In fact, the theory now contains two world-sheet metrics, where
the ``$K$-conjugate'' metric $\tilde{h}_{mn}$ 
(the world-sheet metric of the $\tilde{C}$ theory)
is composed of the
old gauge fields $v,\bar{v}$, and the description of the $K$-conjugate
pair of conformal field theories $C,\tilde{C}$ is now formally symmetric.
To describe the CFT $C$, one views $h_{mn}$ as a fixed world-sheet 
metric and integrates out the spin-two gauge field $\tilde{h}_{mn}$,
and vice-versa to describe the CFT $\tilde{C}$.

The procedure discussed above interprets the $K$-conjugate pair
$C,\tilde{C}$ of CFTs as separate conformal field theories. An
alternative procedure is to integrate out both spin-two gauge
fields $h_{mn}$ and $\tilde{h}_{mn}$, 
which defines a new class of string theories
where the physical states are primary under a $K$-conjugate pair
of Virasoro operators. The first example of this kind of theory was
the ``spin-orbit'' model of \cite{c1} (see also \cite{c7}) and this 
new class of string theories may also be related to the models
of \cite{c18}.


\noindent
{\bf Acknowledgements}

This research is supported in part by
NSF grant PHY-95-14797 and DOE grant DE-AC03-76SF00098.
JdB is a fellow of the Miller Institute for Basic Research in
Science.


\begin{thebibliography}{99}


\small
\parskip=0pt plus 2pt


\bibitem{c1}
K. Bardak\c ci  and M.B. Halpern, {Phys. Rev.}
           {\bf D3} (1971) 2493.
\bibitem{c2}
M.B. Halpern, { Phys. Rev.} {\bf D4} (1971) 2398.       
\bibitem{c3}
P. Goddard, A. Kent and D. Olive,
             { Phys. Lett.} {\bf B152} (1985) 88.               
\bibitem{c4} 
 E. Kiritsis, Mod. Phys. Lett. {\bf A4} (1989) 437.
\bibitem{c5}
M.B. Halpern and E. Kiritsis,
              {Mod. Phys. Lett.} {\bf  A4} (1989) 1373;
             Erratum {\it ibid.} {\bf A4} (1989) 1797.           
\bibitem{c6}
A.Yu Morozov, A.M. Perelomov, A.A. Rosly, M.A. Shifman and
             A.V. Turbiner, {Int. J. Mod. Phys.} {\bf A5} (1990) 803.   
\bibitem{c7}
M.B. Halpern, E. Kiritsis, N.A. Obers and K. Clubok,
  ``Irrational Conformal Field Theory'',
   {Physics Reports} {\bf 265} (1996) 1.                 
\bibitem{c8}
K. Bardak\c{c}i,  E. Rabinovici and B. S\"aring, { Nucl. Phys.}
           {\bf B299} (1988) 151; D. Altschuler, K. Bardak\c{c}i and
           E. Rabinovici, { Comm.  Math. Phys.} { \bf 118} (1988) 241.   
\bibitem{c9}
K. Gawedski and A. Kupainen, Phys.Lett.
{\bf B215} (1988) 119; Nucl. Phys. {\bf B320} (1989) 625.
\bibitem{kahs} D. Karabali, Q-H. Park, H.J. Schnitzer and Z. Yang,
             Phys. Lett. {\bf B216} (1989) 307.
\bibitem{kas} D. Karabali and
             H.J. Schnitzer, Nucl. Phys. {\bf B329} (1990) 649.  
\bibitem{c10}
M.B. Halpern and J.P. Yamron, { Nucl. Phys.} {\bf B351} (1991) 333.      
\bibitem{c11}
J. de Boer, K. Clubok and M.B. Halpern, {Int. J. Mod. Phys.}
             { \bf A9} (1994) 2451.
\bibitem{c12}
K. Clubok and M.B. Halpern, The generic world-sheet
 action of irrational conformal field theory, {\it in}: ``{Strings'95}'',
   I. Bars et al., eds., World Scientific, Singapore (1996).        
\bibitem{c13}
J. de Boer and M.B. Halpern,
      { Int. J. Mod. Phys.}  {\bf A12} (1997) 1551.              
\bibitem{c14}
J. de Boer and M.B. Halpern, ``Unification of the General Non-Linear Sigma
Model and the Virasoro Master Equation'',  
to appear in the Proceedings of the NATO Workshop
``New Developments in Quantum Field Theory'', Zakopane, June 1997, Plenum
Press, N.Y.
\bibitem{c15}
J. de Boer and M.B. Halpern, ``New Spin-Two Gauged Sigma Models and General
Conformal Field Theory'', to appear.
\bibitem{c16}
M.B. Halpern, Recent progress in irrational conformal
             field theory, {\it in}:
             ``{Strings 1993}'',  
M.B. Halpern et al., eds.,  World Scientific,
              Singapore (1995).                              
\bibitem{c17}
M.B. Halpern, E. Kiritsis and N.A. Obers,
             {\it in}: ``{Infinite Analysis}'';
              {Int. J. Mod. Phys.} {\bf A7}, [Suppl.  1A] (1992) 339.    
\bibitem{c18} I. Bars and C. Kounnas, Phys. Rev. {\bf D56} (1997) 3664.


\end{thebibliography}
\end{document}